\documentclass[a4paper]{ifacconf}

\usepackage{natbib}        
\usepackage{amsmath,amssymb,amsfonts}
\usepackage{algorithmic}
\usepackage{graphicx}
\graphicspath{{figure/}}
\newcommand\largefig{0.95}
\usepackage{textcomp}
\usepackage{xcolor}
\usepackage{lipsum}
\usepackage{fixmath}
\usepackage{url}
\usepackage{tikz}
\usetikzlibrary{arrows.meta}
\usetikzlibrary{shapes.misc}
\newtheorem{remark}{Remark}

\newcommand{\RO}{\mathcal{R}_0}
\newcommand{\RB}{\mathcal{R}_B}
\newcommand{\xO}{\mathbold{x_0}}
\newcommand{\yO}{\mathbold{y_0}}
\newcommand{\zO}{\mathbold{z_0}}
\newcommand{\xB}{\mathbold{x_B}}
\newcommand{\yB}{\mathbold{y_B}}
\newcommand{\zB}{\mathbold{z_B}}
\newcommand{\bz}{\mathbold{z}}

\begin{document}
\begin{frontmatter}

\title{{\tiny{\tt Joint IFAC Conference: SSSC, TDS, COSY -- Gif-sur-Vette, France, 30 June - 2 July 2025} \\}
  Guidance and Control of \\
  Unmanned Surface Vehicles via HEOL
}

\author[1]{Lo\"{\i}ck Degorre} %
\author[2]{Emmanuel Delaleau} %
\author[3,6]{C\'edric Join} %
\author[4,5,6]{Michel Fliess}

\address[1]{%
  \textit{ENSTA},
  \textit{UMR 6285},
  \textit{IPP} 
  \textit{Lab-STICC},
  29\,200 Brest, France.
  \texttt{loick.degorre@ensta.fr}%
}

\address[2]{%
  \textit{ENI Brest},
  \textit{UMR CNRS 6027},
  \textit{IRDL},
  29\,200 Brest, France.
  \texttt{emmanuel.delaleau@enib.fr}%
}

\address[3]{%
  \textit{CRAN (CNRS, UMR 7039)},
  \textit{Universit\'{e} de Lorraine},
  \textit{Campus Aiguillettes},
  BP 70239, 54506 Vand{\oe}uvre-l\`{e}s-Nancy, France.
  \texttt{cedric.join@univ-lorraine.fr}%
}

\address[4]{%
  \textit{LIX (CNRS, UMR 7161)},
  \textit{\'{E}cole polytechnique},
  91128 Palaiseau, France.
  \texttt{michel.fliess@polytechnique.edu}%
}
\address[5]{LJLL (CNRS, UMR 7238), Sorbonne Unversit\'e, 75005 Paris, France {michel.fliess@sorbonne-universite.fr}}

\address[6]{%
  \textit{AL.I.E.N.}, 7 rue Maurice Barr\`{e}s,
  54330 V\'{e}zelise, France.\\ %
  \texttt{\{cedric.join, michel.fliess\}@alien-sas.com}%
}

\begin{abstract}%
  This work presents a new approach to the guidance and control of
  marine craft via HEOL, i.e., a new way of combining flatness-based and model-free
  controllers. Its goal is to develop 
  a general %
  regulator for Unmanned Surface Vehicles (USV). %
  To do so, the well-known USV maneuvering model is simplified into a nominal Hovercraft model which is flat. A flatness-based controller is derived for the
  simplified USV model and the loop is closed via an intelligent
  proportional-derivative (iPD) regulator. We thus associate the well-documented natural robustness of
  flatness-based control and adaptivity of iPDs. The controller is
  applied in simulation to two surface vessels, one meeting the
  simplifying hypotheses, the other one being a generic USV of the
  literature. It is shown to stabilize both systems even in the
  presence of unmodeled environmental disturbances.
\end{abstract}


\end{frontmatter}
\section{Introduction}

With the increasing number of applications for both surface and
underwater autonomous vehicles, a great number of control methods and
guidance principles have been developed in recent years, but the need
for efficient and robust trajectory-tracking controllers yet
remains. In this work, surface vessels are considered as a reduced case of
underwater vehicles constrained in the horizontal plane. They are used
as a first step towards the design of all-purpose controllers for
underwater craft.

Control of under-actuated surface vehicles has been addressed in
different manners but mostly for path following and path tracking
applications. See (\cite{PettNijm98cdc,breiviK_principles_2005,elmokadem_trajectory_2016}) for some of the most common methods, and (\cite{degorre_survey_2023}) for a survey.

In the present work,
control of the {\it Unmanned Surface Vehicle} ({\it USV}) is addressed with the HEOL setting, a very interesting candidate for control of marine craft because of its robustness and adaptability. 
In order to use this flatness-based method, a simplified, flat nominal model is extracted from the USV conventional maneuvering model and is used to establish the controller. This simple model is similar to the \textit{Hovercraft} model which is naturally flat (see \cite{sira-ramirez_differentially_2004,Degorre23these}, and the references therein for details).

The so-obtained guidance principle is then applied to standard USVs and, because of the robustness of the method, displays good performance in trajetcory tracking tasks.  



This work illustrates the efficiency of {\em HEOL} (\cite{heol}), which combines in a new manner the well-known \emph{Flatness-Based Control} (Fliess et al. (1995, 1999)) with  \emph{Model-Free Control}~({\em MFC}) (Fliess and Join (2013, 2022)) and the corresponding \emph{intelligent} controllers. Note that HEOL has already been illustrated by \cite{delaleau} and \cite{mfpc}. One should however not forget that 
\begin{itemize}
\item many concrete case-studies have employed MFC since almost twenty years: see, e.g., \cite{gedouin} and \cite{madrid} among a large list of references; 
    \item the connection between flatness and MFC has a long history: see, e.g., \cite{saar} for a recent reference.
\end{itemize}

In HEOL the \emph{ultra-local model} (\cite{FliessJoin13ijc}) is replaced 
by a \emph{homeostat}, which allows the coefficients of the control variables to be obtained immediately. Here an \emph{intelligent proportional-derivative controller} ({\em iPD}) (Fliess and Join (2013, 2022)) is used. The {\em data-driven} term, which helps compensating disturbances and mismatches, is obtained via algebraic estimation techniques (\cite{fliess_algebraic_2003}). 

The paper is organized as follows. Sect.~\ref{sec:models} recalls
the models of both the surface vessel and the hovercraft.
Sect.~\ref{sect:demo_flat} sketches the proof of flatness of the
hovercraft model. Sect.~\ref{sec:controller} develops the iPD.
Sect.~\ref{sec:conclusion} draws some conclusions.

\section{Model of the Vehicle}\label{sec:models}

This section introduces the model of the surface vessel and describes
the simplifying hypotheses leading to the hovercraft model. The models
are derived from the well-known surface vessel
maneuvering model described in \cite{fossen_handbook_2021}.

In this work, the simplified USV, thereafter called ``hovercraft'' for simplicity, is considered to be a surface vessel with a
circular hull-shape and homogeneous mass distribution. The added mass
and damping parameters are then equal in surge and sway. This
hypothesis does simplify the surface vessel model, notably making the
yaw dynamics of the vehicle independent of the surge and sway
dynamics.

Both models are constrained to the horizontal plane and both vehicles
are considered to be actuated with two parallel thrusters generating
independent surge force $\tau_u$ and yaw moment $\tau_r$. They are
then ill-actuated with respect to the positioning tasks at hand, and a
guidance principle is mandatory (\cite{Degorre23these}).

\subsection{Surface vessel model}\label{sect:mod_SV}

To establish the model of a surface marine vessel, one needs to
consider two different frames: the earth-fixed \emph{inertial frame}
$\RO(\mathbold{O_0}, \xO, \yO, \zO)$ and the \emph{body-fixed} frame
$\RB(\mathbold{O_B}, \xB, \yB, \zB)$ centered on the center of gravity
of the vehicle and rotated of an angle $\psi$ around $\zO$
w.r.t.~$\RO$. The variables that appear in the model are: $x,y$ the
coordinates of the vehicle in $\RO$, the translation speed components
$u,v$ in $\RB$, the heading angle of the hovercraft~$\psi$, the
rotation speed $r$ of~$\RB$ w.r.t.~$\RO$. The controls are $F_u$, the
propulsion force, and $\Gamma_r$, the rotating moment normalized in
mass. The surface vessel model is:
\begin{subequations}\label{eq:mod_SV}
    \begin{align}
    \dot x &= u\cos\psi - v\sin\psi \label{eq:mod_SV:x} \\
    \dot y &= u\sin\psi + v\cos\psi \label{eq:mod_SV:y} \\
    \dot\psi &= r \label{eq:mod_SV:psi}\\
    \dot{u} &= F_u + a vr - \beta_u u \label{eq:mod_SV:u} \\
    \dot v &= b ur -  \beta_v v \label{eq:mod_SV:v} \\
    \dot r &= \Gamma_r + cuv- \gamma r \label{eq:mod_SV:r}
    \end{align}
\end{subequations}
the details about the coefficients $a, b, c,\beta_u,\beta_v,\gamma$ are given in the Appendix.

\subsection{Hovercraft model}\label{sect:mod_hov}

\begin{figure}[htbp]
\centerline{\includegraphics[width=\largefig\linewidth,keepaspectratio]{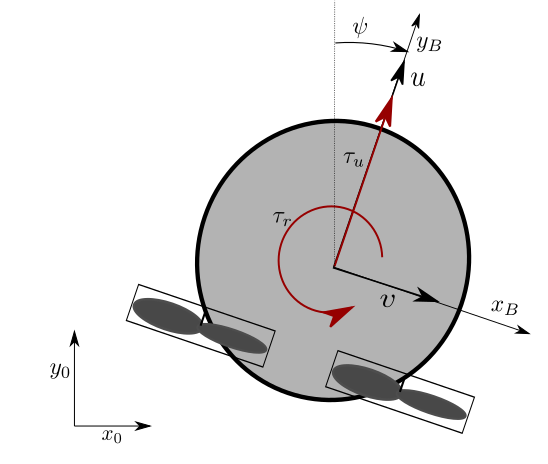}}
\caption{Schematic representation of the hovercraft.}
\label{fig:hov}
\end{figure}

The hovercraft dynamic model is established using the surface vessel model \eqref{eq:mod_SV} and considering a circular hull shape.  The mass ratios $a$, $b$ and $c$ thus become: 
$a = 1$, $b = -1/a = -1$, $c=0 $.
In the same way, the surge and sway damping contributions $\beta_u$ and $\beta_v$ are equivalent. Thus: $ \beta_u = \beta_v = \beta $.

\noindent With these hypotheses, the kinematic part of the model,
\eqref{eq:mod_SV:x} to \eqref{eq:mod_SV:psi},  is not modified, and the surface vessel model
becomes:
\begin{subequations}\label{eq:mod_hov}
    \begin{align}    
    \dot x &= u\cos\psi - v\sin\psi \label{eq:mod_hov:x} \\
    \dot y &= u\sin\psi + v\cos\psi \label{eq:mod_hov:y} \\
    \dot\psi &= r \label{eq:mod_hov:psi}\\
    \dot{u} &= F_u + vr - \beta u \label{eq:mod_hov:u} \\
    \dot v &= - ur -  \beta v \label{eq:mod_hov:v} \\
    \dot r &= \Gamma_r  - \gamma r \label{eq:mod_hov:r}
    \end{align}
\end{subequations}

Note that because of the shape hypothesis, the yaw dynamics described by Eq. \eqref{eq:mod_hov:r} are independent of the surge and sway modes of the vehicle.This assumption is obviously not met in practice on generic surface vessels.


\section{Flatness of the hovercraft model}\label{sect:demo_flat}

\subsection{Preliminary remark}
While the surface vessel model \eqref{eq:mod_SV} is notoriously not flat , the hovercraft
model is flat with the flat output $\bz = [x,y]^\top$ (\cite{sira-ramirez_differentially_2004}). The flatness
proof is briefly recalled here and the reader is referred to
(\cite{Sira2002hovercraft,rigatos_differential_2015,Degorre23these})
for more details.

\subsection{Proof}

Set
\begin{equation}\label{eq:psi_flat}
    \psi = \arctan\left(\frac{\ddot{y} + \beta\dot{y}}{\ddot{x}+ \beta \dot{x}} \right)
\end{equation}
Deriving $\psi$ in Eq. \eqref{eq:psi_flat} yields the yaw velocity $r$ and the normalized yaw moment $\Gamma_r$:  
$$ r = \dot\psi,  \quad \Gamma_r = \ddot\psi +\gamma \dot\psi $$

The heading angle expression~\eqref{eq:psi_flat} unlocks expressions of the surge and sway velocities, as well as the surge force: 
\begin{align}
    u &= \dot{x}\cos\psi + \dot{y}\sin\psi \\
    v &= -\dot{x}\sin\psi + \dot{y}\cos\psi \\
    F_u &= (\ddot{x}+ \beta \dot{x})\cos\psi + (\ddot{y} + \beta\dot{y})\sin\psi \label{eq:tauu_flat}
\end{align}

The hovercraft~\eqref{eq:mod_hov} has been shown to be flat with flat output $\bz = [x,y]^\top$.

\subsection{Brunovsk\'y representation} 

For simplicity, considering the hovercraft model \eqref{eq:mod_hov}, a
first change of input is performed. Because the yaw dynamics described
by \eqref{eq:mod_hov:psi} and~\eqref{eq:mod_hov:r} are decoupled from the
position dynamics, one can proceed to considering the yaw angle $\psi$
as an input of the system instead of the yaw moment $\Gamma_r$. This
change of input is equivalent to assuming that an additional control
stage with much faster converging dynamics ensures that the heading
angle is almost instantaneously stabilized to the reference value
calculated by the flatness-based guidance principle. This is a common
assumption in marine robotics when developing guidance principles
(\cite{mitchell_comparison_2003,breiviK_principles_2005}).

Then, in order to retrieve the Brunovsk\'y-like formulation (\cite{delaleau_control_1998,hagenmeyer_exact_2003}) of the system composed of two independent integrator chains, a second change of input is performed. The system to control thus becomes: 
\begin{subequations}\label{eq:mod_bruno}
\begin{align}
    \dot{x} &= v_x \label{eq:mod_bruno:x}\\
    \dot{v}_x &= w_x  \label{eq:mod_bruno:vx}\\
    \dot{y} &= v_y \label{eq:mod_bruno:y}\\
    \dot{v}_y &= w_y  \label{eq:mod_bruno:vy}
\end{align}
\end{subequations}
with the two new inputs $w_x$ and $w_y$ defined as:
\begin{subequations}
    \begin{align}
        w_x &= F_u\cos\psi - \beta v_x\\
        w_y &= F_u\sin\psi - \beta v_y
    \end{align}
\end{subequations}
The original inputs of the system can be obtained with: 
\begin{subequations}\label{eq:reconstruct}
    \begin{align}
        \psi &= \arctan\left(\frac{w_y + \beta v_y}{w_x + \beta v_x}\right) \label{eq:psi_wxwy}\\
        F_u &= (w_x + \beta v_x)\cos\psi + (w_y + \beta v_y) \sin\psi 
    \end{align}
\end{subequations}

\begin{remark}
    The singularity introduced in \eqref{eq:reconstruct} because of the change of input can be overcome using the well known \emph{arctan2} function and additional usual strategies. Reference trajectories of the task can also be chosen to avoid any singularities. See \cite{degorre_survey_2023} for a precise recall of the \emph{arctan2} function. 
\end{remark}

\begin{figure}[htbp]
    \centering
    \resizebox{.95\linewidth}{!}{%
\begin{tikzpicture}[]
  \filldraw [color=red!70,rounded corners=20mm] (15,20) rectangle (35,40);
  \draw (25,30) node [scale=8] {\textbf{HEOL}};
  \filldraw [color=green!80,rounded corners=20mm] (55,20) rectangle (75,30);
  \draw (65,25) node [scale=8] {\textbf{Autopilot}};
  \filldraw [color=blue!50,rounded corners=20mm] (95,15) rectangle (125,45);
  \draw (110,30) node [scale=8] {\textbf{USV}};
  \draw[](45,25) circle (2);
  \draw[-{Stealth[length=20mm]}] (0,30) -- (15,30);
  \draw (7.5,30) node [scale=10,above] {$x^\ast,y^\ast$};
  \draw[-{Stealth[length=20mm]}] (35,35) -- (95,35);
  \draw (65,35) node [scale=10,above] {$F_u^\ast$};
  \draw[-{Stealth[length=20mm]}] (35,25) -- (43,25);
  \draw (40.0,25) node [scale=10,below] {$\psi  ^\ast$};
  \draw[-{Stealth[length=20mm]}] (47,25) -- (55,25);
  \draw[-{Stealth[length=20mm]}] (75,25) -- (95,25);
  \draw (85,25) node [scale=10,above] {$\Gamma_r$};
  \draw[-{Stealth[length=20mm]}] (125,30) -- (140,30);
  \draw[] (130,30) -- (130,10);
  \draw[] (130,10) -- (45,10);
  \draw[-{Stealth[length=20mm]}] (45,10) -- (45,23);
  \draw (128,10) node [scale=10,above] {$\psi$};
  \draw[] (135,30) -- (135,00);
  \draw[] (135,00) -- (25,00);
  \draw[-{Stealth[length=20mm]}] (25,00) -- (25,20);
  \draw (132,00) node [scale=10,above] {$x,y$};
\end{tikzpicture}
    }
    \caption{Cascade structure}
\label{fig:schema_bloc}
\end{figure}
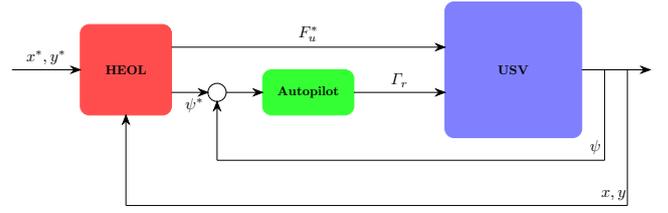

\begin{remark} 
    An additional control stage is necessary to ensure that the heading angle of the vehicle does converge towards the reference value calculated with the guidance principle introduced in the following. As it is often the case with the conventional \emph{Guidance-Control} structure used with autonomous vehicles, the HEOL approach presented in this work leads to a \emph{cascade system}. A standard PID or Sliding Mode controller would be suited to ensure stability of the lower level, provided it is faster than the outer loop. The cascade system is depicted on Fig.~\ref{fig:schema_bloc}. Note also that many autonomous vehicles, especially commercial USVs, feature an embedded autopilot ensuring stability of the lower level of the cascade.
\end{remark}

\section{Design of the controller} \label{sec:controller}

In this section, the controller combining flatness and model-free control is designed. It associates the natural robustness of flatness-based control and exact feedforward linearization \cite{hagenmeyer_robustness_2003} with the adaptivity of Model-Free Control \cite{FliessJoin13ijc}. 
The controller is based on the very simple hovercraft model in the inertial frame \eqref{eq:mod_bruno} and is meant to be applied to the more complex surface vessel. 

The controller is composed of a nominal control calculated in a feedforward linearizing fashion \cite{hagenmeyer_exact_2003} and of an Intelligent Proportional Derivative regulator used to close the loop. The contributions of the nominal controls and the iPDs regulators are linearly summed in the controller. In this case, the robustness and adaptivity of the iPDs allow compensating the mandatory model approximations required for the design of the flatness-based guidance principle.

\subsection{Nominal Controls}

In an exact feedforward linearizing fashion, nominal values of the
controls are computed using the equations of the Brunovsk\'y form
obtained with flatness. Recalling \ref{eq:mod_bruno:vx}
and~\ref{eq:mod_bruno:vy}, one gets the nominal controls:
\begin{subequations}\label{eq:ctrl_nom}
    \begin{align}
        w_x^* &= \dot{v}_x^* = \ddot{x}^* \\
        w_y^* &= \dot{v}_y^* = \ddot{y}^* 
    \end{align}
\end{subequations}
where $x^*$, $y^*$ and their derivatives denote desired values issued
from the trajectory.

\subsection{Closed-loop control design}

Set
\begin{eqnarray*}
    w_x &=& w_x^\ast - \Delta w_x \\
    w_y &=& w_y^\ast - \Delta w_y
\end{eqnarray*}
It yields
    \begin{align*}
        \ddot{e}_x &= \Delta w_x, \quad  e_x = w_x^\ast - w_x \\
        \ddot{e}_y &= \Delta w_y, \quad  e_y = w_y^\ast - w_x
    \end{align*}
and, according to \cite{heol}, the \emph{homeostat}
\begin{equation}\label{homeo}
    \begin{aligned}
        \ddot{e}_x &= F_x + \Delta w_x \\
        \ddot{e}_y &= F_y + \Delta w_y
    \end{aligned}
\end{equation}
where $F_\zeta$, $\zeta = x, y$, accounts for the mismatches and disturbances. Following \cite{mfc2}, an estimate $\widehat F_{\zeta}$ of $F_\zeta$ reads
\bgroup\small
\begin{equation}\label{estimat}
\begin{aligned}
    \widehat F_{\zeta}(t) =
    &\dfrac{60}{T^5} \int_0^T \Big[ \Big((T-\sigma^2) - 4(T-\sigma)\sigma + \sigma^2\Big)e_\zeta(\sigma+t-T) \\
    & \mbox{} + \dfrac{1}{2}(T-\sigma)^2\sigma^2 \Delta w_\zeta(\sigma+t-T)\Big]d\sigma
\end{aligned}
\end{equation}
\egroup
The corresponding \emph{intelligent proportional-derivative controller} ({\em iPD}) reads 
\begin{equation}\label{ipd}
    \Delta w_\zeta = - \left(K_p e_\zeta + K_d\dot{e}_\zeta + \widehat F_\zeta \right)
\end{equation}
where $K_p, K_d \in \mathbb{R}$ are constant gains, such that the polynomial 
$s^2 + K_d s + K_p$ is Hurwitz.

Riachy's trick (\cite{mfc2}) permits to avoid the calculation of the derivative $\dot{e}_\zeta$. Rewrite~\eqref{homeo} as $\ddot{e}_\zeta + K_d \dot{e}_\zeta = F(t) + K_d \dot{e}_\zeta + \Delta w_\zeta$. Set 
$$Y_\zeta(t) = e_\zeta(t) + K_d \int_{c}^t e_\zeta(\sigma)d\sigma, \quad 0 \leqslant c < t$$
It yields $\ddot{Y}_\zeta = \ddot{e}_\zeta + K_d \dot{e}_\zeta$. Set $\mathcal{F}_\zeta = F_\zeta + K_d \dot{e}_\zeta$. Eqn. \eqref{homeo} becomes
    $\ddot{Y}_\zeta = \mathcal{F}_\zeta + \Delta w_\zeta$.
Eqn. \eqref{ipd} reads now
\begin{equation}\label{ipd2}
    \Delta w_\zeta = -(\widehat{\mathcal{F}}_\zeta  +K_p e_\zeta)
\end{equation}
where the derivative of $e_\zeta$ disappears. The estimate $\widehat{\mathcal{F}}$ in Eqn. \eqref{ipd2} may be computed via Formula \eqref{estimat} by replacing $e_\zeta$ by $Y_\zeta$.


Set
\begin{itemize}
\item $K_p^x = K_p^y = K_p$;
\item $K_d^x = K_d^y = K_d$.
\end{itemize}
and only two parameters remain to tune, namely $K_p$ and $K_d$. At last, the inputs of the Brunovsk\'y equivalent system are: 
\begin{subequations}
    \begin{align}
        w_x &= \ddot{x}^\ast + K_d\dot{e}_x + K_p e_x - \widehat F_x \\ 
        w_y &= \ddot{y}^\ast + K_d\dot{e}_y + K_p e_y - \widehat F_y
    \end{align}
\end{subequations}

For added robustness, the original inputs of the simplified hovercraft
system are reconstructed in a feedforward linearizing fashion
(\cite{hagenmeyer_robustness_2003}):
\begin{subequations}\label{eq:reconstruct_final}
    \begin{align}
        \psi &= \arctan\left(\frac{w_y + \beta \dot{x}^*}{w_x + \beta \dot{y}^*}\right) \label{eq:psi_wxwy:star}\\
        F_u &= (w_x + \beta \dot{x}^*)\cos\psi + (w_y + \beta \dot{y}^*) \sin\psi 
    \end{align}
\end{subequations}

The additional control stage used for tracking of the $\psi$ reference calculated by the controller can either be a very simple PID or SMC or a commercial autopilot system.

Overall, the flatness of the simplified hovercraft system allows designing a flatness-based guidance principle, calculating a surge control force and a yaw angle reference out of the position measurements. Then, iPDs are used to close the loop in an adaptive manner, compensating the model approximations of the hovercraft and any unmodeled disturbance.

\section{Simulation Results}
The controller associating flatness and model-free control is tested in simulation in two different scenarios. First, it is tested on a hovercraft in the presence of a constant disturbing force similar to the force of wind. In this case, the algebraic estimators associated to the iPDs compensate the disturbing effect of the wind. 
Then, the controller is evaluated on a surface vessel similar to the \emph{Otter} USV (\emph{Maritime Robotics}, see : \url{https://www.maritimerobotics.com/otter}). This vehicle does not meet the hovercraft simplifying hypothesis. It is evaluated on a circular trajectory in the with the same disturbing external force.

\subsection{Control of the hovercraft}
In this first example, the controller is tested on the hovercraft with an external disturbing wind force. The vehicle has a circular hull and thus $a = 1$, $b=-1$ the normalized damping parameter is chosen as $\beta = 10$. 
The wind is simulated as a constant normalized force in the inertial frame. It is aligned with the $\yO$-axis with a normalized magnitude equal to $-50$. 

The hovercraft is tested on a simple straight segment defined as $x^*(t)=2t$, $y^*(t)=0$. It is initiated with a $10$m initial error on the $\yO$ axis. 

\begin{figure}[htbp]
\centerline{\includegraphics[width=\largefig\linewidth,keepaspectratio]{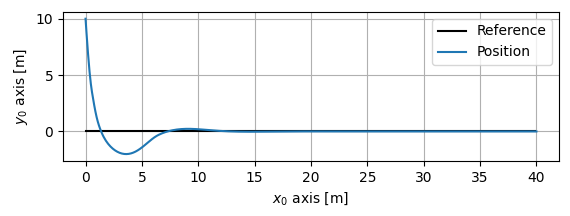}}
\caption{Trajectory of the hovercraft in the $(\xO, \yO$) plane, in presence of a wind force aligned with the $\yO$ axis. Black: Reference trajectory - Blue: Actual trajectory of the vehicle}
\label{fig:traj:hov}
\end{figure}

\begin{figure}[htbp]
\centerline{\includegraphics[width=\largefig\linewidth,keepaspectratio]{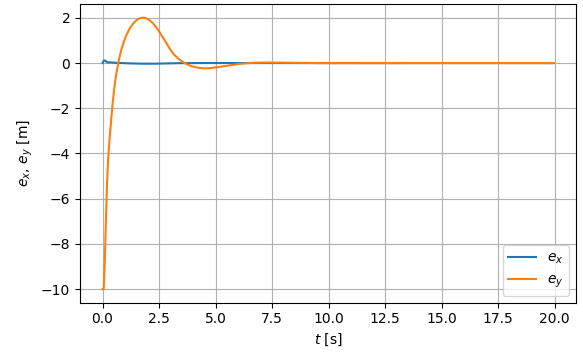}}
\caption{Tracking errors of the hovercraft over time. Blue: Error on the $\xO$ axis - Orange: Error on the $\yO$ axis}
\label{fig:err:hov}
\end{figure}

Figs.~\ref{fig:traj:hov} and~\ref{fig:err:hov} display the behavior of
the vehicle and of the tracking errors on this task. Even with the
considerable initial error, both errors converge to zero rapidly.

\begin{figure}[htbp]
\centerline{\includegraphics[width=\largefig\linewidth,keepaspectratio]{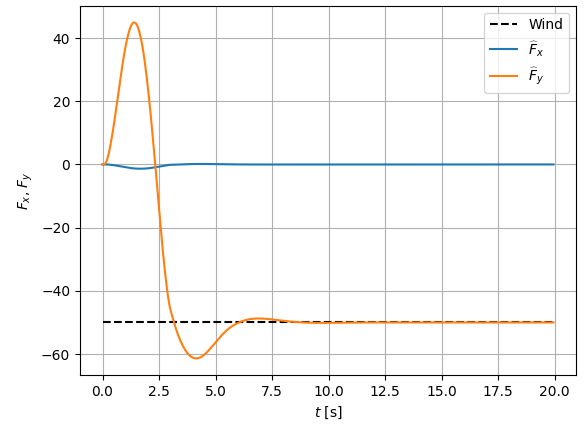}}
\caption{Estimated values of $F_x$ and $F_y$ over time for the hovercraft. Dashed Black: Wind force - Blue: $\widehat{F}_x$ - Orange:$\widehat{F}_y$}
\label{fig:esti:hov}
\end{figure}

Fig. \ref{fig:esti:hov} shows that the $\yO$-axis estimator $F_y$ does converge towards the disturbing force while $F_x$ stays around zero. This figure confirms that without model approximations, the estimators of the iPDs behave like the integral terms of conventional PIDs and compensate the unmodeled disturbance.

This first example shows that, on the very simple task of tracking line with a hovercraft, the controller behaves as expected. The estimators compensate all disturbing effects and the allow tracking without steady state error. 

\subsection{Control of a standard USV}
In this second example, the controller is applied to a USV similar to the \emph{Otter} USV (the parameters are derived from \cite{fossen_mss}). On this vehicle the parameters are: 
\begin{align*}
    a = 0.58, \quad b = -1.72, \quad \beta = \beta_u = 10, \quad \beta_v = 1.5\beta
\end{align*}
The vehicle does not meet the hypothesis of a circular hull shape. Because the surge and sway damping parameters are different, one of them must be chosen for the controller calculations. In this case, the smaller one $\beta_u$ is chosen. 

The controller is evaluated on a circular trajectory with the same $\yO$-axis disturbing force as before. The circular trajectory has been chosen to showcase the behavior of the vehicle for all orientations and demonstrate that the singularity due to the arctangent function is of no consequence. The vehicle's position is initialized with a $15$m error on the $\xO$ axis.

\begin{figure}[htbp]
\centerline{\includegraphics[width=\largefig\linewidth,keepaspectratio]{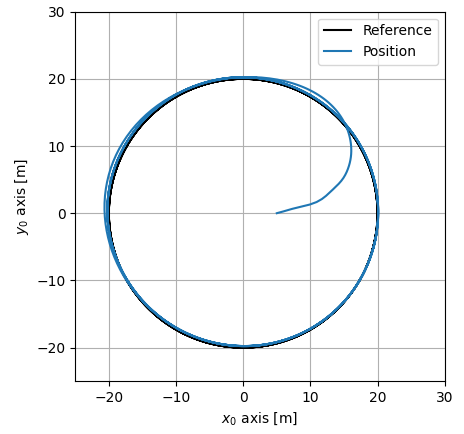}}
\caption{Trajectory of the surface vessel in the $(\xO, \yO$) plane, in presence of a wind force aligned with the $\yO$ axis. Black: Reference trajectory - Blue: Actual trajectory of the vehicle}
\label{fig:traj:sv}
\end{figure}

\begin{figure}[htbp]
\centerline{\includegraphics[width=\largefig\linewidth,keepaspectratio]{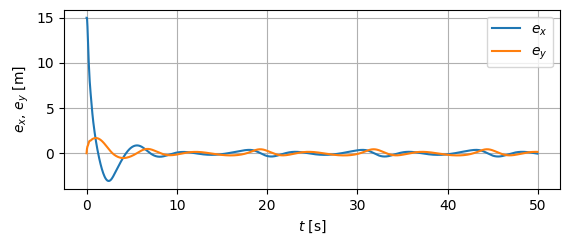}}
\caption{Tracking errors of the surface vessel over time. Blue: Error on the $\xO$ axis - Orange: Error on the $\yO$ axis}
\label{fig:err:sv}
\end{figure}

Figs.~\ref{fig:traj:sv} and~\ref{fig:err:sv} show that even with the consequent model
approximation due to the shape of the surface vessel used in this
example, the controller ensures convergence of the position of the
vehicle towards the desired trajectory. Both errors are maintained
very close to zero during all the application.

\begin{figure}[htbp]
\centerline{\includegraphics[width=\largefig\linewidth,keepaspectratio]{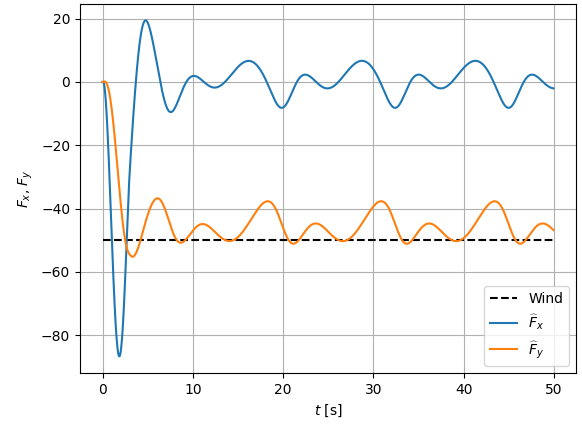}}
\caption{Estimated values of $F_x$ and $F_y$ over time for the surface vessel. Dashed Black: Wind force - Blue: $\widehat{F}_x$ - Orange:$\widehat{F}_y$}
\label{fig:esti:sv}
\end{figure}

Fig.~\ref{fig:esti:sv} depicts the behavior of the two estimators on the
surface vessel example. Both oscillate around the expected values:
$\widehat F_x$ stays around zero and $\widehat F_y$ converges towards
the wind value and oscillates around it. The oscillations observed on
Fig.~\ref{fig:esti:sv} are expected and demonstrate that the
estimators of the iPDs also counteract the model approximations. In
facts, one can observe two phase-shifted sinusoidal signals on the
estimators curves. They correspond to the disturbances relative to the
speed of the vehicle due to the error on the damping coefficient and
the acceleration due to the approximation of the mass parameters.

Overall, our HEOL control setting, which associates a flatness-based guidance
principle and a model-free regulator, shows very good performances on
the circle trajectory when applied to this surface vessel even though
it was calculated using the model of a hovercraft (compare with \cite{Sira2002hovercraft} and \cite{rigatos_differential_2015})

\section{Conclusion} \label{sec:conclusion}

This work presents a novel approach to guidance and control of autonomous marine craft, via HEOL, which combines flatness-based and model-free controls. Considering the hovercraft vehicle as a drastic simplification of the well-known surface vessel maneuvering model, a flatness-based guidance principle is established. The flatness of the hovercraft model gives equations of the surge force generated by the two parallel thrusters and of the heading angle of the vehicle that can be used as a guidance principle. Then, two decoupled model-free controllers relying on Intelligent PDs are used to close the loop. They add an additional layer of robustness to the controller, ensuring compensation of all external disturbing effects as well as compensating the model approximations due to usage of the hovercraft model. The controller has then been applied in simulation to a hovercraft and to a surface vessel with generic hull-shape and shows very good trajectory tracking results in both cases even in the presence of external disturbances. This work hints towards an all-purpose controller for marine craft based on the new promising HEOL standpoint. 












\begin{thebibliography}{21}
\providecommand{\natexlab}[1]{#1}
\providecommand{\url}[1]{\texttt{#1}}
\providecommand{\urlprefix}{URL }
\expandafter\ifx\csname urlstyle\endcsname\relax
  \providecommand{\doi}[1]{doi:\discretionary{}{}{}#1}\else
  \providecommand{\doi}{doi:\discretionary{}{}{}\begingroup
  \urlstyle{rm}\Url}\fi

\bibitem[Artuñedo et al.(2024)]{madrid}
Artuñedo, A., Moreno-Gonzalez, M., and Villagra, J. (2024).
\newblock Lateral control for autonomous vehicles: A comparative evaluation.
\newblock {\em Annual Rev. Contr.}, 57, 100910.

\bibitem[{Breivik and Fossen(2005)}]{breiviK_principles_2005}
Breivik, M. and Fossen, T. (2005).
\newblock Principles of {guidance}-{based} {path} {following} in {2D} and {3D}.
\newblock \emph{{IEEE} {Conf.} {Dec.} {Contr.}}, 627--634, Seville.

\bibitem[{Degorre(2023)}]{Degorre23these}
Degorre, L. (2023).
\newblock \emph{Analysis and Control of Autonomous Underwater Vehicles with
Reconfigurable Vectoring Thrust}.
\newblock Ph.D. Thesis, \'Ecole Nat. Ing\'en. Brest.

\bibitem[{Degorre et~al.(2023)Degorre, Delaleau, and
  Chocron}]{degorre_survey_2023}
Degorre, L., Delaleau, E., and Chocron, O. (2023).
\newblock A survey on model-based control and guidance principles for
  autonomous marine vehicles.
\newblock \emph{J. Marine Sci. Engin.}, 11, 430.

\bibitem[Delaleau et~al.(2025)]{delaleau}
Delaleau, E., Join, C., and Fliess, M. (2025).
\newblock Synchronization of Kuramoto oscillators via HEOL, and a  discussion on AI.
\newblock \emph{IFAC PapersOnLine}, 59-1, 229--234.

\bibitem[{Delaleau and Rudolph(1998)}]{delaleau_control_1998}
Delaleau, E., and Rudolph, J. (1998).
\newblock Control of flat systems by quasi-static feedback of generalized
  states.
\newblock \emph{Int. J. Contr.}, 71, 745--765.

\bibitem[{Elmokadem et~al.(2016)Elmokadem, Zribi, and
  Youcef-Toumi}]{elmokadem_trajectory_2016}
Elmokadem, T., Zribi, M., and Youcef-Toumi, K. (2016).
\newblock Trajectory tracking sliding mode control of underactuated {AUVs}.
\newblock \emph{Nonlin. Dyn}, 84, 1079--1091.

\bibitem[{Fliess and Join(2013)}]{FliessJoin13ijc}
Fliess, M. and Join, C. (2013).
\newblock Model-free control.
\newblock \emph{Int. J. Contr.}, 86, 2228--2252.

\bibitem[Fliess and Join(2022)]{mfc2}
Fliess, M., and Join, C. (2022). An alternative to proportional-integral and proportional-integral-derivative regulators: Intelligent proportional-derivative regulators. {\em Int. J. Robust Nonlin. Contr.}, 32, 9512--9524.

\bibitem[Fliess et al. (1995)]{flmr_ijc}
Fliess, M., L{\'e}vine, J., Martin, P., and Rouchon, P. (1995). Flatness and defect of non-linear systems: introductory theory and examples. {\em Int. J. Contr.}, 61, 1327--1361.

\bibitem[Fliess et al.(1999)]{flmr_ieee}
Fliess, M., L\'{e}vine, J., Martin, P., and Rouchon, P. (1999).
A Lie-B\"acklund approach to equivalence and flatness
of nonlinear systems. {\em IEEE Trans. Autom. Contr.}, 44, 922--937.

\bibitem[{Fliess and Sira-Ram\'{\i}rez(2003)}]{fliess_algebraic_2003}
Fliess, M. and Sira-Ram\'{\i}rez, H. (2003).
\newblock An algebraic framework for linear identification.
\newblock \emph{ESAIM: COCV}, 9, 151--168.

\bibitem[{Fossen(2021)}]{fossen_handbook_2021}
Fossen, T.I. (2021).
\newblock \emph{Handbook of Marine Craft Hydrodynamics and Motion Control} (2nd ed.).
\newblock Wiley.

\bibitem[{Fossen and Perez(2004)}]{fossen_mss}
Fossen, T.I., and Perez, T. (2004).
\newblock Marine systems simulator ({MSS}) - {M}atlab {T}oolbox.

\bibitem[G\'{e}douin et al.(2011)]{gedouin}
G\'{e}douin, P.-A., Delaleau, E., Bourgeot, J.-M., Join, C., Arbab Chirani, S., and Calloch, S. (2011).
\newblock Experimental comparison of classical PID and model-free control: Position control of a shape memory alloy active spring.
\newblock \emph{Contr. Engin. Pract.}, 19, 433--441.

\bibitem[{Hagenmeyer and Delaleau(2003{\natexlab{a}})}]{hagenmeyer_exact_2003}
Hagenmeyer, V., and Delaleau, E. (2003{\natexlab{a}}).
\newblock Exact feedforward linearization based on differential flatness.
\newblock \emph{International Journal of Control}, 76(6), 537--556.

\bibitem[{Hagenmeyer and
  Delaleau(2003{\natexlab{b}})}]{hagenmeyer_robustness_2003}
Hagenmeyer, V., and Delaleau, E. (2003{\natexlab{b}}).
\newblock Robustness analysis of exact feedforward linearization based on
  differential flatness.
\newblock \emph{Automatica}, 39, 1941--1946.

\bibitem[Join et al.(2024)]{heol}
Join, C., Delaleau, E., and Fliess, M. (2024). Flatness-based control revisited: The {\it HEOL} setting. \emph{C.R. Math.}, 362, 1693--1706.

\bibitem[Join et al.(2025)]{mfpc}
Join, C., Delaleau, E., Fliess, M. (2025).
\newblock Model-free predictive control: Introductory algebraic calculations, and a brief comparison with HEOL. \emph{This Conference}. {\tt arXiv:2502.00443} 


\bibitem[{Mitchell et~al.(2003)Mitchell, McGookin, and
  Murray-Smith}]{mitchell_comparison_2003}
Mitchell, A., McGookin, E., and Murray-Smith, D. (2003).
\newblock Comparison of {Control} {Methods} for {Autonomous} {Underwater}
  {Vehicles}.
\newblock \emph{IFAC Proc. Volumes}, 36, 37--42.



\bibitem[{Pettersen and Nijmeijer(1998)}]{PettNijm98cdc}
Pettersen, K. and Nijmeijer, H. (1998).
\newblock Tracking control of an underactuated surface vessel.
\newblock In \emph{IEEE Conference on Decision and Control}, volume~4,
  4561--4566.


\bibitem[{Rigatos(2015)}]{rigatos_differential_2015}
Rigatos, G.G. (2015).
\newblock Differential {flatness} {theory} and {flatness}-{based} {control}.
\newblock In G.G. Rigatos (ed.), \emph{Nonlinear {Control} and {Filtering}
{Using} {Differential} {Flatness} {Approaches}: {Applications} to
{Electromechanical} {Systems}}. Springer.

\bibitem[{Scherer et~al.(2023)Scherer, Othmane, and
Rudolph}]{saar}
Scherer, P., Othmane, A., and Rudolph, J. (2023).
\newblock Combining model-based and model-free approaches for the control of an
electro-hydraulic system.
\newblock \emph{Contr. Engin. Pract.}, 133, 105453.


\bibitem[{Sira-Ram\'{\i}rez(2002)}]{Sira2002hovercraft}
Sira-Ram\'{\i}rez, H. (2002).
\newblock Dynamic second-order sliding mode control of the hovercraft vessel.
\newblock \emph{IEEE Trans. Control Sys. Tech.}, 10, 860--865.

\bibitem[{Sira-Ram\'{\i}rez and
Agrawal(2004)}]{sira-ramirez_differentially_2004}
Sira-Ram\'{\i}rez, H. and Agrawal, S.K. (2004).
\newblock \emph{Differentially flat systems}.
\newblock Marcel Dekker.


\end{thebibliography}

\appendix
\section{}\label{sec:coefficients}
The elements of this Section follows~\cite{fossen_handbook_2021}. The parameters $m$, $I_z$, $X_{\dot{u}}$, $Y_{\dot{v}}$ and $N_{\dot{r}}$ are respectively the mass, the moment of inertia, the surge, sway, and yaw added masses. Only linear decoupled damping is considered, $d_u$, $d_v$ and $d_r$ are the surge, sway, and yaw damping coefficients respectively. Finally, reduced parameters $a$, $b$, $c$, $\beta_u$, $\beta_v$ and $\gamma$ are introduced for ease of explanation. Moreover, $\tau_u$ is the surge force generated by the thrusters, $\tau_r$ is the yaw moment and $F_u$ and $\Gamma_r$ act as normalized inputs of the surface vessel model. The expressions of the parameters of model~\eqref{eq:mod_SV} are:
\begin{align*}
    a &= \frac{m-Y_{\dot{v}}}{m-X_{\dot{u}}}, \quad b = -\frac{1}{a}, \quad c = \frac{X_{\dot{u}}-Y_{\dot{v}}}{I_z-N_{\dot r}} \\
    \beta_u &= \frac{d_u}{m-X_{\dot u}}, \quad \beta_v = \frac{d_v}{m-Y_{\dot{v}}} \quad \gamma = \frac{d_r}{{I_z-N_{\dot r}}} \\
    F_u &= \frac{\tau_u}{m-X_{\dot u}} \quad \Gamma_r = \frac{\tau_r }{I_z-N_{\dot r}}
\end{align*}

The hovercraft has a circular hull shape, consequently, the added mass parameters of the hovercraft are then equal in surge and sway: $X_{\dot{u}} = Y_{\dot{v}}$.  

\end{document}